\documentclass[12pt]{article}
\usepackage{amsmath,amssymb,bm,epsfig}
\usepackage[subrefformat=parens]{subcaption}
\usepackage{cite}
\usepackage{here,color}

\renewcommand{\thefootnote}{\fnsymbol{footnote}}

\begin{document}

\title{
\begin{flushright}
\begin{minipage}{0.2\linewidth}
\normalsize
EPHOU-18-002
\\*[50pt]
\end{minipage}
\end{flushright}
{\Large \bf {
Neutrino mixing from finite modular groups
\\*[20pt]}}}

\author{
Tatsuo~Kobayashi\footnote{
E-mail address: kobayashi@particle.sci.hokudai.ac.jp
}, \ \
Kentaro~Tanaka\footnote{
E-mail address: k-tanaka@particle.sci.hokudai.ac.jp
}, and \
Takuya~H.~Tatsuishi\footnote{
E-mail address: t-h-tatsuishi@particle.sci.hokudai.ac.jp
}\\*[20pt]
{\it \normalsize 
Department of Physics, Hokkaido University, Sapporo 060-0810, Japan} \\*[50pt]}

\date{
\centerline{\small \bf Abstract}
\begin{minipage}{0.9\linewidth}
\medskip 
\medskip 
\small
We study the lepton flavor models, whose flavor symmetries are finite subgroups of the modular group such as $S_3$ and $A_4$.
In our models, couplings are also nontrivial representations of these groups and modular functions of the modulus.
We study the possibilities that these models realize realistic values of neutrino masses and lepton mixing angles.
\end{minipage}
}

\begin{titlepage}
\maketitle
\thispagestyle{empty}
\clearpage
\tableofcontents
\thispagestyle{empty}
\end{titlepage}

\renewcommand{\thefootnote}{\arabic{footnote}}
\setcounter{footnote}{0}
\vspace{35pt}
\section{Introduction}

One of the unsolved but important mysteries in particle physics is the mystery about 
the flavor structure of the quarks and leptons such as 
the generation number, mass hierarchy and mixing angles.
Such a mystery would provide us with hints to explore physics beyond 
the standard model.
Indeed, several types of scenarios have been proposed to understand 
quark and lepton masses and mixing angles as well as CP phases.

One of interesting ideas is to impose non-Abelian discrete flavor symmetries.
Many models have been proposed imposing non-Abelain discrete flavor symmetries, 
e.g., $S_3$, $A_4$, $S_4$ and other various finite groups.
(See for review \cite{Altarelli:2010gt,Ishimori:2010au,King:2013eh,King:2015aea}.)
In particular, the lepton sector has been intensively studied, 
because at least two of three lepton mixing angles are large compared with the quark mixing angles 
and their experimental results have been improved precisely.
Recently, CP symmetry and its violation were also studied \cite{Feruglio:2012cw,Holthausen:2012dk,Chen:2014tpa}.

Superstring theory is a promising candidate for the unified theory of 
all interactions including gravity and matter fields such as 
quarks and leptons as well as Higgs fields.
It is shown that superstring theory on certain compactifications 
leads to non-Abelian discrete flavor symmetries.
For example, heterotic string theory on orbifolds can lead to
$D_4$, $\Delta(54)$, etc. \cite{Kobayashi:2006wq}.
(See also \cite{Kobayashi:2004ya,Ko:2007dz}.)\footnote{
In Ref.\cite{Beye:2014nxa}, relations between enhanced gauge symmetries 
and non-Abelian discrete flavor symmetries were studied.} 
In addition type II magnetized and intersecting D-brane models can lead to 
similar flavor symmetries \cite{Abe:2009vi,Abe:2009uz,BerasaluceGonzalez:2012vb,Marchesano:2013ega,Abe:2014nla}.

On the other hand, string theory on tori and orbifolds has the modular symmetry.
For example, modular symmetries were studied in heterotic orbifold models \cite{Lauer:1989ax,Lerche:1989cs,Ferrara:1989qb} 
and magnetized D-brane models 
\cite{Cremades:2004wa,Kobayashi:2017dyu,Kobayashi:2018rad}. 
In general, modular transformations act nontrivially on string modes and interchange 
massless modes such as quarks and leptons to each other.
In this sense, modular symmetry is a non-Abelain discrete flavor symmetry.
Furthermore, it is interesting that  the modular symmetry includes $S_3$, $A_4$, $S_4$, $A_5$ 
as its congruence subgroups, $\Gamma(N)$.
However, there is a difference between the modular symmetry and the usual flavor symmetries.\footnote{
See for their difference and relation \cite{Kobayashi:2018rad}.}
Coupling constants such as Yukawa couplings also transform nontrivially under the modular symmetry  
\cite{Lauer:1989ax,Cremades:2004wa,Kobayashi:2017dyu,Kobayashi:2018rad}, 
while coupling constants are invariant under the usual flavor symmetries, 
although flavon fields, which develop their vacuum expectation values (VEVs), transform nontrivially 
under flavor symmetries.
Moreover, Yukawa couplings as well as higher order coupling constants are modular functions of 
moduli \cite{Hamidi:1986vh,Cvetic:2003ch,Cremades:2004wa,Abe:2009dr}.

By use of the above aspects, 
an interesting ansatz was proposed in Ref. \cite{Feruglio:2017spp}, 
where $\Gamma(3) \simeq A_4$ was used and leptons were assigned to triplets and singlets of $A_4$.
Also coupling constants were assigned to $A_4$ triplet and singlets, which are 
modular functions.
Then, neutrino masses and mixing angles were analyzed.
Such an ansatz would be interesting in order to bridge a gap between underlying theory such as superstring theory 
and low-energy physics like neutrino phenomena.
Our purpose in this paper is to study systematically the above approach.
We study the $A_4$ model following Ref. \cite{Feruglio:2017spp}.
Also, we extend our analysis to $S_3$ models.

This paper is organized as follows.
In section \ref{sec:1}, we give a briel review on 
modular symmetry.
In section \ref{sec:data}, using experimental values, 
we write the neutrino mass matrix, which is convenient to 
our analyses.
In section \ref{sec:A4}, we study systematically the $A_4$ model 
 following Ref. \cite{Feruglio:2017spp}.
 In section \ref{sec:S3},  we construct $S_3$ models in a way 
 similar to the $A_4$ model and study them systematically.
 Section \ref{sec:Conclusion} is our conclusion and discussions.
In appendix \ref{sec:app-1}, we briefly review modular functions and 
show modular functions corresponding to the $A_4$ triplet and the $S_3$ doublet.

\section{Modular transformation and its subgroup symmetries}
\label{sec:1}

In this section, we give a brief review on the modular symmetry and its 
congruence subgroups.

Toroidal compactification is one of simple compactifications.
The two-dimensional torus $T^2$ is obtained as $T^2 = {\mathbb R}^2/\Lambda$.
Here, $\Lambda$ is the two-dimensional lattice, which is spanned by 
two lattice vectors, $\alpha_1=2\pi R$ and $\alpha_2 = 2\pi R \tau$.
We use the complex coordinate, where $R$ is a real parameter, 
and $\tau$ is a complex modulus parameter.

The choice of the basis vectors has some ambiguity.
The same lattice can be spanned by other bases,
\begin{equation}
\label{eq:SL2Z}
\left(
\begin{array}{c}
\alpha'_2 \\ \alpha'_1
\end{array}
  \right) =\left(
  \begin{array}{cc}
a & b \\
c & d   
\end{array}
\right) \left(
  \begin{array}{c}
\alpha_2 \\ \alpha_1
\end{array}
  \right) ,
\end{equation}
where $a,b,c,d$ are integer with satisfying $ad-bc = 1$.
That is the $SL(2,Z)$ transformation.

The modular parameter $\tau = \alpha_2/\alpha_1$ transforms as 
\begin{equation}\label{eq:tau-SL2Z}
\tau \longrightarrow \tau'= \frac{a\tau + b}{c \tau + d},
\end{equation}
under (\ref{eq:SL2Z}).
Both lattice bases $(\alpha_1,\alpha_2)$ and $(\alpha'_1,\alpha'_2)$, and both 
modular parameters, $\tau$ and $\tau'$, lead to the same lattice $\Lambda$ and the same $T^2$.
The modular transformation is generated by the $S$ and $T$ transformations, 
\begin{eqnarray}
& &S:\tau \longrightarrow -\frac{1}{\tau}, \\
& &T:\tau \longrightarrow \tau + 1.
\end{eqnarray}
In addition, these satisfy the following algebraic relations,
\begin{equation}
S^2 =1, \qquad (ST)^3 =1.
\end{equation}
On top of that, when we impose $T^N = 1$, 
the so-called  congruence subgroups $\Gamma(N)$ can be 
realized.
The congruence subgroups $\Gamma(N)$ are isomorphic to 
(even) permutation groups, e.g. 
$\Gamma(2) \simeq S_3$, $\Gamma(3) \simeq A_4$, 
$\Gamma(4) \simeq S_4$, and $\Gamma(5) \simeq A_5$.
(For subgroups of the modular group, e.g., see Ref.~\cite{deAdelhartToorop:2011re}.)

String theory on $T^2$ as well as orbifolds $T^2/Z_N$ has the modular symmetry.
Furthermore, four-dimensional low-energy effective field theory on the 
compactification $T^2\times X_4$ as well as $(T^2/Z_N)\times X_4$ 
also has the modular symmetry, where 
$X_4$ is a four-dimensional compact space.

A set of chiral superfields $\phi^{(I)}$ transform under the modular transformation (\ref{eq:tau-SL2Z}) as 
a multiplet  \cite{Ferrara:1989bc},
\begin{equation}
\phi^{(I)}\to(c\tau+d)^{-k_I}\rho^{(I)}(\gamma)\phi^{(I)},
\end{equation}
where  $-k_I$ is the so-called modular weight and $\rho^{(I)}$ denotes a representation matrix.
Modular invariant kinetic terms expanded around a VEV of the modulus $\tau$ are written by
\begin{equation}
 \frac{|\partial_\mu\tau|^2}{\langle\tau-\bar{\tau}\rangle^2}+\sum_I\frac{|\partial_\mu\phi^{(I)}|^2}{\langle\tau-\bar{\tau}\rangle^{k_I}}.
\end{equation}
Also, the superpotential should be invariant under the modular symmetry.
That is, the superpotential should have vanishing modular weight in global supersymmetric models.
Indeed, Yukawa coupling constants as well as higher-order couplings 
constants are modular functions of $\tau$ 
\cite{Hamidi:1986vh,Cvetic:2003ch,Cremades:2004wa,Abe:2009dr}.
In the framework of supergravity theory, the superpotential must be invariant up to the K\"ahler transformation \cite{Ferrara:1989bc}.
That implies that the superpotential of supergravity models with the above kinetic term should have 
modular weight one.
In sections \ref{sec:A4} and \ref{sec:S3}, we consider the global supersymmetric models, 
and require that the superpotential has vanishing modular weight, 
although it is straightforward to arrange modular weights of chiral superfields for supergravity models.

The Dedekind eta-function $\eta(\tau)$ is one of famous modular functions, which is written by 
\begin{equation}
\eta(\tau) = q^{1/24} \prod_{n =1}^\infty (1-q^n),
\end{equation}
where $q = e^{2 \pi i \tau}$.
The $\eta(\tau)$ function behaves under $S$ and $T$ transformations as 
\begin{equation}\label{eq:eta-ST}
\eta(-1/\tau) = \sqrt{-i \tau}\eta(\tau), \qquad \eta(\tau + 1) = e^{i\pi/12}\eta(\tau).
\end{equation}
The former transformation implies that the $\eta(\tau)$ function has the modular weight $1/2$.

The modular functions $(Y_1,Y_2,Y_3)$ with weight 2, which behave as 
an $A_4$ triplet, are obtained as 
\begin{eqnarray} 
\label{eq:Y-A4}
Y_1(\tau) &=& \frac{i}{2\pi}\left( \frac{\eta'(\tau/3)}{\eta(\tau/3)}  +\frac{\eta'((\tau +1)/3)}{\eta((\tau+1)/3)}  
+\frac{\eta'((\tau +2)/3)}{\eta((\tau+2)/3)} - \frac{27\eta'(3\tau)}{\eta(3\tau)}  \right), \nonumber \\
Y_2(\tau) &=& \frac{-i}{\pi}\left( \frac{\eta'(\tau/3)}{\eta(\tau/3)}  +\omega^2\frac{\eta'((\tau +1)/3)}{\eta((\tau+1)/3)}  
+\omega \frac{\eta'((\tau +2)/3)}{\eta((\tau+2)/3)}  \right) , \\
Y_3(\tau) &=& \frac{-i}{\pi}\left( \frac{\eta'(\tau/3)}{\eta(\tau/3)}  +\omega\frac{\eta'((\tau +1)/3)}{\eta((\tau+1)/3)}  
+\omega^2 \frac{\eta'((\tau +2)/3)}{\eta((\tau+2)/3)}  \right) , \nonumber
\end{eqnarray}
in Ref. \cite{Feruglio:2017spp}, where 
$\omega= e^{2\pi i/3}$.
(See Appendix \ref{sec:app-1}.)

We can obtain the modular functions with weight 2, which behave as 
an $S_3$ doublet,
\begin{eqnarray} 
Y_1(\tau) &=& \frac{i}{4\pi}\left( \frac{\eta'(\tau/2)}{\eta(\tau/2)}  +\frac{\eta'((\tau +1)/2)}{\eta((\tau+1)/2)}  
- \frac{8\eta'(2\tau)}{\eta(2\tau)}  \right) ,\nonumber \\
Y_2(\tau) &=& \frac{\sqrt{3}i}{4\pi}\left( \frac{\eta'(\tau/2)}{\eta(\tau/2)}  -\frac{\eta'((\tau +1)/2)}{\eta((\tau+1)/2)}   \right)  ,
 \label{eq:modular function_S3 doublet}
\end{eqnarray}
by a similar technique.
(See Appendix \ref{sec:app-1}.)

We use the following expansions:
\begin{eqnarray} 
Y_1(\tau) &=& \frac 18 + 3q + 3q^2 + 12 q^3 + 3q^4 \cdots ,\nonumber \\
Y_2(\tau) &=& \sqrt 3 q^{1/2} (1+ 4 q + 6 q^{2} + 8 q^{3} \cdots  ).  \label{eq:modular func_S3_expand}
\end{eqnarray}
In section \ref{sec:S3}, we fit $Y_2/Y_1$ to experimental data.
That is, we use the following expansion:
\begin{equation}
\frac{1}{Y_1} = \frac18 \cdot \frac{1}{1+ 24 q + \cdots}.
\end{equation}
Such an expansion would be valid for $|24q| \lesssim 0.1$, i.e. 
\begin{equation}
\label{eq:Im-tau}
\text{Im} (\tau) \gtrsim 0.65.
\end{equation}

\section{Neutrino mass matrix}
\label{sec:data}

Before studying the $A_4$ and $S_3$ models, here 
using experimental values of neutrino oscillations
we write the neutrino mass matrix,  which is convenient to our analyses in 
Sections \ref{sec:A4} and \ref{sec:S3}.

\subsection{Experimental values}

Flavor eigenstates of neutrino $(\nu_e,\nu_\mu,\nu_\tau)$ are linear combinations of mass eigenstates $(\nu_1,\nu_2,\nu_3)$.
Their mixing matrix $U$, i.e. the so-called PMNS matrix can be written by
\begin{align}
 U&=
\begin{pmatrix}
 U_{e1} & U_{e2} & U_{e3} \\
 U_{\mu1} & U_{\mu2} & U_{\mu3} \\
 U_{\tau1} & U_{\tau2} & U_{\tau3} \\
\end{pmatrix} \notag \\
 &=
 \begin{pmatrix}
 1 & 0 & 0 \\
 0 & c_{23} & s_{23} \\
 0 & -s_{23} & c_{23} \\
 \end{pmatrix}
 \begin{pmatrix}
 c_{13} & 0 & s_{13}e^{-i\delta_{CP}} \\
 0 & 1 & 0 \\
 -s_{13}e^{i\delta_{CP}} & 0 & c_{13} \\
 \end{pmatrix}
 \begin{pmatrix}
 c_{12} & s_{12} & 0 \\
 -s_{12} & c_{12} & 0 \\
 0 & 0 & 1 \\
 \end{pmatrix}
 \begin{pmatrix}
 1 & 0 & 0 \\
 0 & e^{i\alpha_2/2} & 0 \\
 0 & 0 & e^{i\alpha_3/2} \\
 \end{pmatrix},
\end{align}
where $c_{ij}=\cos\theta_{ij}$ and $s_{ij}=\sin\theta_{ij}$ for mixing angles $\theta_{ij}$, $\delta_{CP}$ is the Dirac CP phase, and $\alpha_i$ are Majorana CP phases.
The mass-squared differences are defined by
\begin{align}
 &\delta m^2=m_2^2-m_1^2, \\
 &\Delta m^2=m_3^2-\frac{m_1^2+m_2^2}{2},
\end{align}
where $m_i$ is the mass eigenvalue of $\nu_i$.
We also define the ratio between the mass-squared  differences as
\begin{equation}
 r=\frac{\delta m^2}{|\Delta m^2|}.
\end{equation}
Experimental values with normal ordering (NO) and inverted ordering (IO) are shown in Table~\ref{tab:experimental values}.
\begin{table}[htbp]
 \centering
\begin{equation*}
\begin{array}{|c|c|c|} \hline
 \text{Parameter} & \text{Normal Ordering} & \text{Inverted Ordering} \\ \hline
 \delta m^2/10^{-5} \text{eV}^2 & 7.37^{+0.17}_{-0.16} & 7.37^{+0.17}_{-0.16} \\ \hline
 |\Delta m^2|/10^{-3}\text{eV}^2 & 2.525^{+0.042}_{-0.030}  & 2.505^{+0.034}_{-0.032} \\ \hline
 \sin^2\theta_{12}/10^{-1} & 2.97^{+0.17}_{-0.16}  & 2.97^{+0.17}_{-0.16} \\ \hline
 \sin^2\theta_{13}/10^{-2} & 2.15^{+0.07}_{-0.07}  & 2.16^{+0.08}_{-0.07} \\ \hline
 \sin^2\theta_{23}/10^{-1} & 4.25^{+0.21}_{-0.15}  & 5.89^{+0.16}_{-0.22}\oplus 4.33^{+0.15}_{-0.16} \\ \hline
 \delta_{CP}/\pi & 1.38^{+0.23}_{-0.20}  & 1.31^{+0.31}_{-0.19} \\ \hline
 r & 2.92^{+0.10}_{-0.11}\times10^{-2} & 2.94^{+0.11}_{-0.10}\times10^{-2} \\ \hline
\end{array}
\end{equation*}
 \caption{The best-fit values and $1\sigma$-ranges in experiments with NO and IO from Ref.~\cite{Capozzi:2017ipn}.
}
 \label{tab:experimental values}
\end{table}

\subsection{General setup of our models}

In sections \ref{sec:A4} and \ref{sec:S3}, we consider the models, where the charged lepton mass matrix is 
diagonal, following \cite{Feruglio:2017spp}.
In such models, the neutrino mass matrix $m_\nu$ is written as 
\begin{equation}
m_\nu=U^\ast\text{diag}(m_1,m_2,m_3)U^\dagger ,
\end{equation}
by use of the PMNS matrix.
By using three column vectors $U=(\vec{U}_1,\vec{U}_2,\vec{U}_3)$, the matrix $m_\nu$ can be decomposed into
\begin{equation}
m_\nu = m_\nu^{(1)}+e^{-i\alpha_2}m_\nu^{(2)}+e^{-i\alpha_3}m_\nu^{(3)},
\end{equation}
where
\begin{equation}
m_\nu^{(1)} \equiv m_1(\vec{U}_1^\ast\cdot\vec{U}_1^\dagger), \qquad 
e^{-i\alpha_2}m_\nu^{(2)} \equiv m_2(\vec{U}_2^\ast\cdot\vec{U}_2^\dagger), \qquad 
e^{-i\alpha_3}m_\nu^{(3)}  \equiv m_3(\vec{U}_3^\ast\cdot\vec{U}_3^\dagger) .
\end{equation}
The matrices $m_\nu^{(i)}$ are symmetric matrices and 
do not depend on Majorana phases $\alpha_i$ in this definition.
The overall sizes of $m_\nu^{(i)}$ are of ${\cal O}(m_i)$.

In the case of NO ($m_1\simeq m_2\ll m_3$), $m_\nu$ can be approximated by
\begin{equation}
m_\nu\simeq m_\nu^{(3)}=m_3
 \begin{pmatrix}
 s_{13}^2e^{2i\delta_{CP}} & c_{13}s_{13}s_{23}e^{i\delta_{CP}} & c_{13}s_{13}c_{23}e^{i\delta_{CP}} \\
 & c_{13}^2s_{23}^2 & c_{13}^2c_{23}s_{23} \\
 && c_{13}^2c_{23}^2
 \end{pmatrix},
\end{equation}
up to the phase.
Here, we have omitted to write explicitly some elements, because 
$m^{(3)}_\nu$ is the symmetric matrix.
These matrix elements have the following relations:
\begin{equation}
\begin{split}
 &(m_\nu)_{11}(m_\nu)_{22}=(m_\nu)_{12}^2,\quad(m_\nu)_{11}(m_\nu)_{33}=(m_\nu)_{13}^2,\quad(m_\nu)_{22}(m_\nu)_{33}=(m_\nu)_{23}^2, \\
 &\frac{(m_\nu)_{22}}{(m_\nu)_{33}}=t_{23}^2,\quad \frac{(m_\nu)_{11}}{(m_\nu)_{22}+(m_\nu)_{33}}=t_{13}^2e^{2i\delta_{CP}},
\end{split}
\label{eq:consistency conditions_NO}
\end{equation}
where $t_{ij}=\tan\theta_{ij}$.
Thus, realistic models should be consistent with these rules (\ref{eq:consistency conditions_NO}).

In the case of IO ($m_3\ll m_1\simeq m_2$), the mass matrix $m_\nu$ can be approximated by
\begin{equation}
 m_\nu\simeq m_\nu^{(1)}+e^{-i\alpha_{2}}m_\nu^{(2)} \label{eq:mass matrix_IO} .
\end{equation}
Since $\alpha_2$ has not been determined by experiments, there is one more parameter in IO than in NO.

\section{$A_4$ model}
\label{sec:A4}

In this section, we consider the model in Ref.~\cite{Feruglio:2017spp} systematically.
This model is the supersymmetric model, although we can construct a similar nonsupersymmetric model.
This model has the flavor symmetry $\Gamma(3)\simeq A_4$.

\begin{table}[htbp]
 \centering
\begin{equation*}
\begin{array}{|c|c|c|c|} \hline
  & SU(2)_L \times U(1)_Y & A_4 & k_I \\ \hline
 e_{R_1}^c & ({\bf 1},+1) & {\bf 1} & k_{e1} \\ \hline
 e_{R_2}^c & ({\bf 1},+1) & {\bf 1}^{\prime\prime} & k_{e2} \\ \hline
 e_{R_3}^c & ({\bf 1},+1) & {\bf 1}^\prime & k_{e3} \\ \hline
 L & ({\bf 2},-1/2) & {\bf 3} & k_L \\ \hline
 H_u & ({\bf 2},+1/2) & {\bf 1} & k_{H_u} \\ \hline
 H_d & ({\bf 2},-1/2) & {\bf 1} & k_{H_d} \\ \hline
 \phi & ({\bf 1},0) & {\bf 3} & k_\phi \\ \hline
\end{array}
\end{equation*}
 \caption{$A_4$ representations and $k_I$ in the $A_4$ model}
 \label{tab:charge assignment_A4}
\end{table}
We concentrate on the lepton sector.
Table ~\ref{tab:charge assignment_A4} shows the $A_4$ representations and $k_I$ of lepton and Higgs superfields, $L_i$, $e^c_{Ri}$ and $H_{u,d}$.
Recall that $-k_I$ is the modular weight.
The superfield $\phi$ is a flavon field and the $A_4$ triplet.
We arrange $k_I$ such that the charged lepton masses are not modular functions of $\tau$ and 
the flavon field does not appear in the Weinberg operator.
For example, we take $k_\phi=3$, $k_{H_{u,d}}=0$, $k_{L}=1$ and $k_{ei}=-4$ \cite{Feruglio:2017spp}.
Then, the superpotential terms of the charged lepton sector can be written by,
\begin{equation}
W_e =
\beta_1 e_{R_1}^cH_d(L\phi)_1+\beta_2 e_{R_2}^cH_d(L\phi)_{1^\prime}+\beta_3 e_{R_3}^cH_d(L\phi)_{1^{\prime\prime}},
\end{equation}
where the $\beta_i$ are constant coefficients.

We assume that the flavon multiplet develops the VEV along the direction, $\langle\phi\rangle=(u,0,0)$.
Such a VEV as well as the VEV $v_d$ of the neutral component of $H_d$ leads to the diagonal charged lepton mass matrix,
\begin{equation}
m_e = uv_d\left(
\begin{array}{ccc}
\beta_1 & & \\
& \beta_2 & \\
& & \beta_3
\end{array}\right).
\end{equation}
By choosing proper values of couplings $\beta_i$, we can realize the experimental values of 
the charged lepton masses, $m_{e,\mu,\tau}$.

On the other hand, we can write the Weinberg operator in the superpotential
\begin{equation}
 W_\nu=\frac{1}{\Lambda}\left(H_uH_uLLY(\tau)\right)_1. \label{eq:superpotential_A4_nutrino}
\end{equation}
Since $k_L=1$ and $k_{H_u}=0$, 
 the couplings  $Y=(Y_1,Y_2,Y_3)$ must be a modular form with modular weight~2 and an $A_4$ triplet.
 We use the modular functions (\ref{eq:Y-A4}).
Then, we obtain the mass matrix of neutrinos written by
\begin{equation}
 m_\nu^\text{model}=\frac{v^2_u}{\Lambda}
\begin{pmatrix}
 2Y_1 & -Y_3 & -Y_2 \\
 -Y_3 & 2Y_2 & -Y_1 \\
 -Y_2 & -Y_1 & 2Y_3 \\
\end{pmatrix},
\label{eq:massmatrix_A4}
\end{equation}
where $v_u$ denotes the VEV of the neutral component of $H_u$.

Note that the charged lepton mass matrix is diagonal.
Thus, mixing angles as well as the CP phases originated from the neutrino mass matrix.
In the following subsections, we perform numerical analyses by using the mass matrix (\ref{eq:massmatrix_A4}).

\subsection{Normal ordering in $A_4$ model}

Here, we study the NO case.
In the case of NO, there are constraints (\ref{eq:consistency conditions_NO}) for the realistic mass matrix.
By using the equations in the second line of (\ref{eq:consistency conditions_NO}), we obtain 
\begin{equation}
 t_{13}^2e^{2i\delta_{CP}}=\frac{Y_1}{Y_2+Y_3}=\left[4\left(1+\frac{Y_2}{Y_3}\right)\frac{Y_2}{Y_3}\right]^{-1}=\frac{1}{4t_{23}^2(1+t_{23}^2)}.
\label{eq:consistency condition_NO_A4}
\end{equation}
This is the theoretical prediction in this model.
By putting experimental values with $\pm3\sigma$, the left- and right-hand sides of (\ref{eq:consistency condition_NO_A4}) become
\begin{equation}
 \left|t_{13}^2e^{2i\delta_{CP}}\right|\sim0.02,\quad \frac{1}{4t_{23}^2(1+t_{23}^2)}\sim0.2,
\label{eq:consistency condition_NO_A4_ex}
\end{equation}
respectively.
The theoretical prediction (\ref{eq:consistency condition_NO_A4}) is inconsistent with experimental data (\ref{eq:consistency condition_NO_A4_ex}).
Thus, this model does not reproduce the experimental results for NO.

\subsection{Inverted ordering in $A_4$ model}

Here, we study the IO case.
The mass matrix in this model $m_\nu^\text{model}$ (\ref{eq:massmatrix_A4}) obeys the following relations:
\begin{align}
 &F_1(\alpha_2;s_{ij},\delta_{CP})\equiv (m_\nu^\text{model})_{11}+2(m_\nu^\text{model})_{23}=0, \\
 &F_2(\alpha_2;s_{ij},\delta_{CP})\equiv (m_\nu^\text{model})_{22}+2(m_\nu^\text{model})_{13}=0, \\
 &F_3(\alpha_2;s_{ij},\delta_{CP})\equiv (m_\nu^\text{model})_{33}+2(m_\nu^\text{model})_{12}=0.
\end{align}

\begin{figure}[htbp]
 \centering
 \includegraphics[width=70mm]{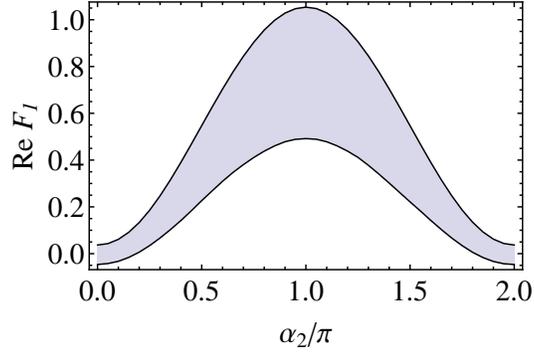}
 \caption{Plot of $\text{Re}\,F_1$ with $3\sigma$ deviation of $s_{ij}$ and $\delta_{CP}$}
 \label{fig:A4Y1}
\end{figure}

\begin{figure}[htbp]
 \centering
 \begin{tabular}{c}
  \begin{minipage}{0.5\hsize}
   \centering
   \includegraphics[width=70mm]{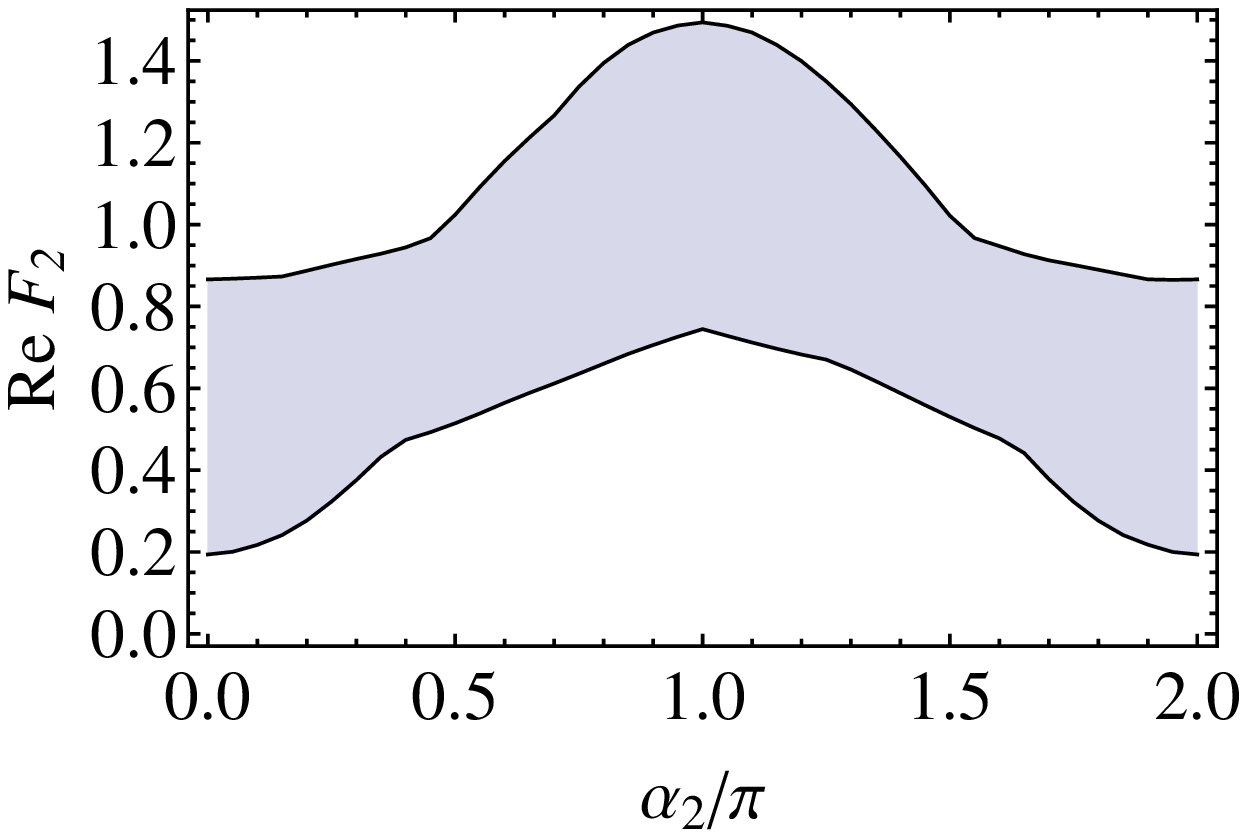}
  \end{minipage}
  \begin{minipage}{0.5\hsize}
   \centering
   \includegraphics[width=70mm]{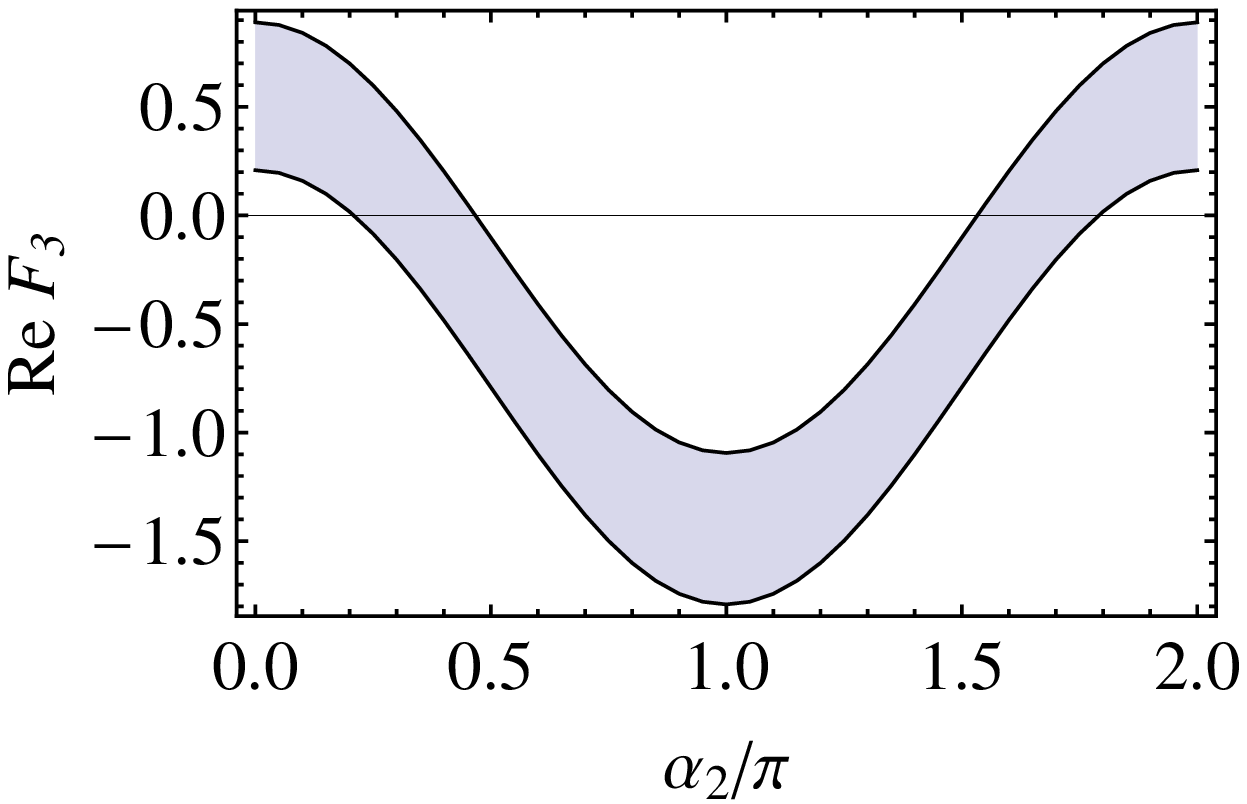}
  \end{minipage}
 \end{tabular}
 \caption{Plots of $\text{Re}\,F_2$ (left) and $\text{Re}\,F_3$ (right) with $3\sigma$ deviation of $s_{ij}$ and $\delta_{CP}$}
 \label{fig:A4Y2Y3}
\end{figure}

Figure \ref{fig:A4Y1} shows a plot of $\text{Re}\,F_1$.
The shaded region shows the $3\sigma$ deviation of $s_{ij}$ and $\delta_{CP}$.
Figure \ref{fig:A4Y2Y3} shows plots of $\text{Re}\,F_i,\,i=2,3$.
Since a realistic model must satisfy $F_i=0$ at the same $\alpha_2$, this model does not reproduce the experimental results within the $3\sigma$ range for IO.

\section{$S_3$ model}
\label{sec:S3}

In this section, we construct the models with the flavor symmetry  $\Gamma(2)\simeq S_3$ in a way 
similar to the $A_4$ model, and study them systematically.

\begin{table}[htbp]
 \centering
\begin{equation*}
\begin{array}{|c|c|c|c|} \hline
  & SU(2)_L \times U(1)_Y & S_3 & k_I \\ \hline
 e_{R_a}^c & ({\bf 1},+1) & {\bf 1} & -3 \\ \hline
 e_{R_b}^c & ({\bf 1},+1) & {\bf 1} & -4 \\ \hline
 e_{R_c}^c & ({\bf 1},+1) & {\bf 1}^\prime & -4 \\ \hline
 L^{(1)} & ({\bf 2},-1/2) & {\bf 1} & 1\\ \hline
 L^{(2)} & ({\bf 2},-1/2) & {\bf 2} & 1 \\ \hline
 H_u & ({\bf 2},+1/2) & {\bf 1} & 0 \\ \hline
 H_d & ({\bf 2},-1/2) & {\bf 1} & 0 \\ \hline
 \phi^{(1)} & ({\bf 1},0) & {\bf 1} & 2 \\ \hline
 \phi^{(2)} & ({\bf 1},0) & {\bf 2} & 3 \\ \hline
\end{array}
\end{equation*}
 \caption{ $S_3$ representations and $k_I$ in the $S_3$ models.}
 \label{tab:charge assignment_S3}
\end{table}
Table~\ref{tab:charge assignment_S3}  shows the $S_3$ representations and $k_I$ of lepton and Higgs superfields.
The $\phi^{(1)}$ and $\phi^{(2)}$ fields are flavon fields, and $\phi^{(1)}$ and $\phi^{(2)}$ are $S_3$ singlet and doublet, 
respectively.
In order to distinguish $e_{R_a}$ and $e_{R_b}$, we assign $k_I$ different from each other.
For such a purpose, we can impose an additional symmetry, e.g. $Z_2$.
We assign $k_I$ such that we can realize the diagonal charged lepton mass matrix similar to the $A_4$ model.
Indeed, the superpotential terms in the charged lepton sector can be written by 
\begin{equation}
 W_e=
\beta_a e_{R_a}^cH_d(L^{(1)}\phi^{(1)})_1+\beta_b e_{R_b}^cH_d(L^{(2)}\phi^{(2)})_1-\beta_c e_{R_c}^cH_d(L^{(2)}\phi^{(2)})_{1^\prime},
\label{eq:charged electron_S3}
\end{equation}
where the $\beta_i$ are constant coefficients.
We assume that the flavon fields develop their VEVs as 
\begin{equation}
 \langle \phi^{(1)}\rangle=u_1,\quad\langle \phi^{(2)}\rangle=(u_2,0).
\end{equation}
Then, we can realize the diagonal charged lepton mass matrix when the neutral component of $H_d$ develops its VEV.
Similar to the $A_4$ model, we can realize the experimental values of the charged lepton masses, $m_{e,\mu,\tau}$ 
by choosing proper values of couplings $\beta_a$.
Note that the  assignment of generations to $e_{R_i},\,i=a,b,c$ is not fixed yet.

Modular invariant Weinberg operators in the superpotential can be written by
\begin{align}
 \mathcal{L}_\text{eff}^\nu
 &=\frac{1}{\Lambda}\left[dHH\left(L^{(2)}L^{(2)}\right)_2Y^{(2)}+aHH\left(L^{(1)}L^{(2)}\right)_2Y^{(2)}\right. \notag \\
 &\quad\quad\left.+bHH\left(L^{(1)}L^{(1)}\right)_1Y^{(1)}+cHH\left(L^{(2)}L^{(2)}\right)_1Y^{(1)}\right], \label{eq:Lagrangian_S3}
\end{align}
where $a,b,c,d\in\mathbb{C}$ are constant coefficients.
$Y^{(1)}$ and $Y^{(2)}$ are modular forms with modular weight 2, and 
$Y^{(1)}$ and $Y^{(2)}$ are $S_3$ singlet and doublet
\footnote{
There are two independent modular forms with weight 2 and $\Gamma(2)$ \cite{Feruglio:2017spp,Gunning:1962}.
Thus, there is only one independent modular form doublet $Y^{(2)}$ in (\ref{eq:Lagrangian_S3}).
}
, respectively.
Note that since $\textbf{1}'$ in $\textbf{2}\times\textbf{2}=\textbf{1}+\textbf{1}'+\textbf{2}$ is antisymmetric, $\left(L^{(2)}L^{(2)}\right)_{1'}=0$.
We denote $Y^{(1)}=Y$ and $Y^{(2)}=(Y_1,Y_2)$.
There are 6 ways to assign 3 generations of lepton $L_i$ to $S_3$ singlet $L^{(1)}$ and doublet $L^{(2)}$.
The replacement of $L^{(2)}=(L_i,L_j)\to(L_j,L_i)$ corresponds to the replacement of $Y_1\leftrightarrow Y_2$ and less affects analysis.
Therefore, we study 3 models shown in table (\ref{tab:S3 models}).
\begin{table}[htbp]
 \centering
\begin{equation*}
\begin{array}{|c|c|c|c|} \hline
 \text{Model} & L^{(1)} & L^{(2)} & m_\nu^\text{model} \\ \hline
 1 & L_3 & (L_1,L_2) & 
 \begin{pmatrix}
  dY_2&dY_1&aY_1 \\
  dY_1&-dY_2&aY_2\\
  aY_1&aY_2&0
 \end{pmatrix}
+
 \begin{pmatrix}
  cY&0&0 \\
  0&cY&0\\
  0&0&bY
 \end{pmatrix} \\ \hline
 2 & L_2 & (L_1,L_3) & 
 \begin{pmatrix}
      dY_2&aY_1&dY_1 \\
      aY_1&0&aY_2\\
      dY_1&aY_2&-dY_2
 \end{pmatrix}
+
 \begin{pmatrix}
      cY&0&0\\
      0&bY&0\\
      0&0&cY
 \end{pmatrix} \\ \hline
 3 & L_1 & (L_2,L_3) & 
 \begin{pmatrix}
      0&aY_1&aY_2 \\
      aY_1&-dY_2&dY_1\\
      aY_2&dY_1&dY_2
 \end{pmatrix}
+
 \begin{pmatrix}
     bY&0&0\\
     0&cY&0\\
     0&0&cY
 \end{pmatrix} \\ \hline
\end{array}
\end{equation*}
 \caption{Three $S_3$ models}
 \label{tab:S3 models}
\end{table}
The fifth column shows the neutrino mass matrix in each model.
Since parameters $b$ and $c$ always appear in $bY$ and $cY$, respectively, we rewrite $B\equiv bY$ and $C\equiv cY$.

\subsection{Normal ordering in $S_3$ models}

Here, we study the NO case in the three $S_3$ models.
In the model~1, the neutrino mass matrix is written by 
\begin{equation}
m_\nu^\text{model}=
 \begin{pmatrix}
  C+dY_2&dY_1&aY_1 \\
  dY_1&C-dY_2&aY_2\\
  aY_1&aY_2&B
 \end{pmatrix}.
\end{equation}
Since the charged lepton mass matrix is diagonal, the PMNS matrix is determined only by the neutrino mass 
matrix.
The consistency conditions (\ref{eq:consistency conditions_NO}) are written by 
\begin{align}
 &(C+dY_2)(C-dY_2)=(dY_1)^2,\quad B(C+dY_2)=(aY_1)^2,\quad B(C-dY_2)=(aY_2)^2, \label{eq:consistency_S3_NO_model1_1}  \\
 &\frac{C-dY_2}{B}=t_{23}^2,\quad \frac{C+dY_2}{B+(C-dY_2)}=e^{2i\delta_{CP}}t_{13}^2. \label{eq:consistency_S3_NO_model1_2}
\end{align}
We obtain $dY_2=C/2$ from (\ref{eq:consistency_S3_NO_model1_1}), and then
\begin{equation}
 t_{23}^2=\frac{C}{2B},\quad e^{2i\delta_{CP}}t_{13}^2=\frac{3}{1+\frac{2B}{C}}
\end{equation}
from (\ref{eq:consistency_S3_NO_model1_2}).
Thus, a prediction of this model is
\begin{equation}
 e^{2i\delta_{CP}}t_{13}^2=3s_{23}^2 \label{eq:consistency_S3_NO_model1_3}.
\end{equation}
Experimental values lead to $t_{13}^2=\mathcal{O}(10^{-2})$ and $3s_{23}^2=\mathcal{O}(1)$.
Hence this model is inconsistent with experiments.
Note that since this result (\ref{eq:consistency_S3_NO_model1_3}) does not depend on $Y_i$, we obtain the same result by replacing $Y_1\leftrightarrow Y_2$, that is, $L^{(2)}=(L_1,L_2)\to(L_2,L_1)$.

In the models~2 and 3, we can analyze in the same way as in the model 1.
A prediction of the model~2 is
\begin{equation}
 e^{2i\delta_{CP}}t_{13}^2=3c_{23}^2,
\end{equation}
and a prediction of the model~3 is
\begin{equation}
 t_{23}^2=3.
\end{equation}
Both of these predictions are inconsistent with experiments.
Hence, these models are not realistic.

\subsection{Inverted ordering in $S_3$ models}

Here, we study the IO case in the three $S_3$ models.

\subsubsection{Model 1}
In the model~1, the mass matrix is written by 
\begin{equation}
m_\nu^\text{model}=
 \begin{pmatrix}
  C+dY_2&dY_1&aY_1 \\
  dY_1&C-dY_2&aY_2\\
  aY_1&aY_2&B
 \end{pmatrix}.
\end{equation}
From this matrix, we find 
\begin{equation}
 \frac{(m_\nu^\text{model})_{11}-(m_\nu^\text{model})_{22}}{2(m_\nu^\text{model})_{12}}
=\frac{(m_\nu^\text{model})_{23}}{(m_\nu^\text{model})_{13}}=\frac{Y_2}{Y_1},
\end{equation}
and hence
\begin{equation}
 F^\text{model}\equiv 2(m_\nu^\text{model})_{12}(m_\nu^\text{model})_{23}-(m_\nu^\text{model})_{13}\left[(m_\nu^\text{model})_{11}-(m_\nu^\text{model})_{22}\right]=0.
\label{eq:consistency_S3_IO_model1}
\end{equation}
This is a prediction of this model.

First, we define a function
\begin{equation}
 F_1(\alpha_2;s_{ij},\delta_{CP})=2(m_\nu)_{12}(m_\nu)_{23}-(m_\nu)_{13}\left[(m_\nu)_{11}-(m_\nu)_{22}\right],
\end{equation}
with $m_\nu$ in (\ref{eq:mass matrix_IO}), and search a set of the values of $(\alpha_2,s_{ij},\delta_{CP})$ satisfying $F_1=0$ within the $3\sigma$ experimental range.
Here, we treat $\alpha_2$ as a free parameter.
\begin{figure}[htbp]
 \centering
 \includegraphics[width=70mm]{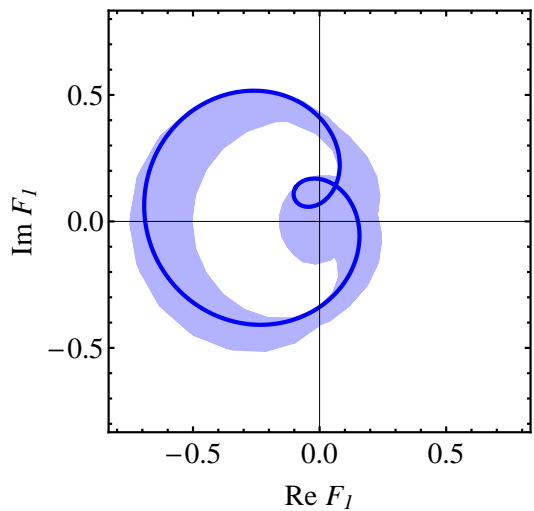}
 \caption{$F_1$ in the model 1.}
 \label{fig:F4cp}
\end{figure}
Figure \ref{fig:F4cp} shows a plot of the $F_1$ on a complex plane.
The blue curve is an one-parametric plot of the $F_1(\alpha_2)$ with $\alpha_2\in[0,2\pi)$ and the best-fit values of $s_{ij}$ and $\delta_{CP}$.
The blue shaded region is a two-parametric plot of the $F_1(\alpha_2;\delta_{CP})$ with $\alpha\in[0,2\pi)$, $\delta_{CP}\in(3\sigma \text{-range})$, and the best-fit values of $s_{ij}$.
The origin $F_1=0$ is the point consistent with the model prediction (\ref{eq:consistency_S3_IO_model1}), and this point corresponds to $\alpha_2/\pi\simeq0.26$ and $\delta_{CP}/\pi\simeq0.89$.

Second, we solve $m_\nu^\text{model}=m_\nu$ with numerical matrix $m_\nu$.
Since there are six complex model parameters and six complex equations, we can solve these equations.
An approximate solution is
\begin{equation}
 \begin{split}
  &a=0.874-0.162i,\quad B=0.469-0.246i,\quad C=0.607-0.235i, \\
  &d=-0.473-0.269i,\quad Y_1=0.163+0.330i,\quad Y_2=-0.468+0.218i.
 \end{split}
\label{eq:solution_S3_model1}
\end{equation}
At this stage, numerically expressed mass matrix $m_\nu$ has degenerate eigenvalues $m_1^2=m_2^2$, 
and therefore $m_\nu$ does not reproduce the mixing angle $s_{12}^2$ and mass-squared  differences.

We can resolve the degeneracy  $m_1^2=m_2^2$ by changing parameters slightly.
That is, we modify the parameter $B\to B+\epsilon_B$ to reproduce the $s_{12}^2$ and mass-squared  differences by resolving the degeneracy.
When the $\epsilon_B=0.030+0.026i$, observables are given by
\begin{equation}
 \begin{split}
 &s_{12}^2=2.79\times10^{-1},\quad s_{13}^2=2.37\times10^{-2},\quad s_{23}^2=5.95\times10^{-1}, \\
 &r=2.91\times10^{-2},\quad\sin\delta_{CP}=-0.58,
 \end{split}
\end{equation}
and Majorana phases are given by
\begin{equation}
 \alpha_2/\pi=0.39,\quad\alpha_3/\pi=0.70.
\end{equation}
These observables are consistent with experimental results up to $3\sigma$, and the values of Majorana phases are predictions of the model 1.

Finally, we fit the values of $Y_1$ and $Y_2$ in (\ref{eq:solution_S3_model1}) with the modular functions in (\ref{eq:modular function_S3 doublet}).
Since the overall coefficient of $Y_i(\tau)$ is not fixed in (\ref{eq:modular func_S3_doublet}), we fit the value of $Y_2/Y_1$ by
\begin{equation}
 \frac{Y_2(\tau)}{Y_1(\tau)}\simeq1.403\times e^{0.507\pi i}.
\end{equation}
Such a ratio can be obtained for  $\tau=0.505+0.781i$.
This value is sufficiently large compared with Eq.~(\ref{eq:Im-tau}).

When we change the generation assignment as $L^{(2)}=(L_1,L_2)\to(L_2,L_1)$, $Y_1$ and $Y_2$ are replaced each other.
Thus, we fit the values of $Y_1$ and $Y_2$ by
\begin{equation}
 \frac{Y_2(\tau)}{Y_1(\tau)}\simeq0.713\times e^{-0.507\pi i},
\end{equation}
and we get the solution $\tau=-0.507+0.960i$.

\subsubsection{Model 2}
In the model~2, we can analyze with the same way as in the model~1.

\begin{figure}[htbp]
 \centering
 \begin{tabular}{c}
  \begin{minipage}{0.5\hsize}
   \centering
   \includegraphics[width=70mm]{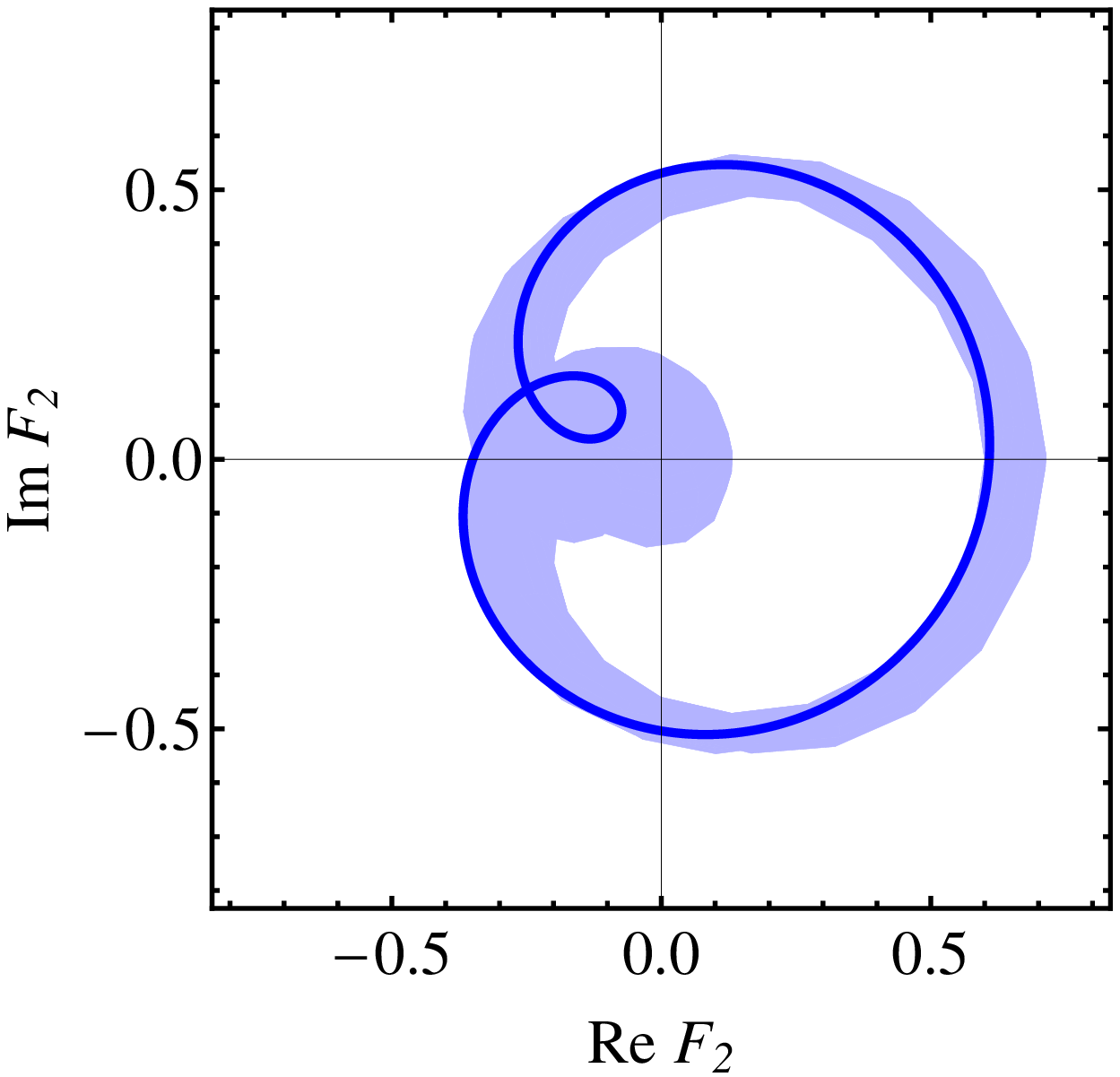}
 \caption{$F_2$ in the model 2}
 \label{fig:F5cp}
  \end{minipage}
  \begin{minipage}{0.5\hsize}
   \centering
   \includegraphics[width=70mm]{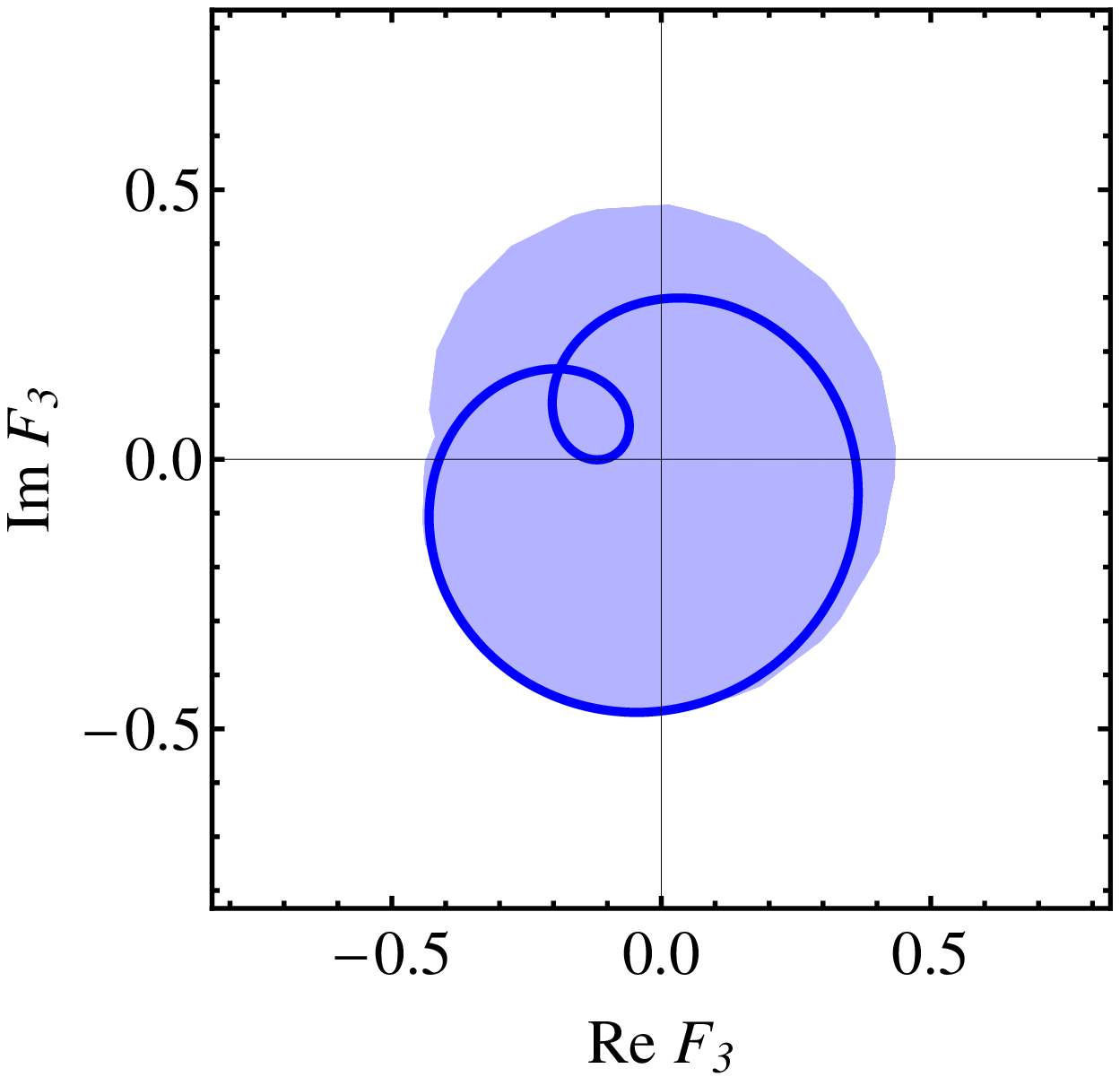}
 \caption{$F_3$ in the model 3}
 \label{fig:F6cp}
  \end{minipage}
 \end{tabular}
\end{figure}

The model~2 predicts 
\begin{equation}
 F^\text{model}\equiv 2(m_\nu^\text{model})_{13}(m_\nu^\text{model})_{23}-(m_\nu^\text{model})_{12}\left[(m_\nu^\text{model})_{11}-(m_\nu^\text{model})_{33}\right]=0.
\label{eq:consistency_S3_IO_model2}
\end{equation}
We define
\begin{equation}
 F_2(\alpha_2;s_{ij},\delta_{CP})=2(m_\nu)_{13}(m_\nu)_{23}-(m_\nu)_{12}\left[(m_\nu)_{11}-(m_\nu)_{33}\right],
\end{equation}
and this is shown in Figure \ref{fig:F5cp} similar to the model~1.
The condition $F_2=0$ is satisfied at $\alpha_2/\pi\simeq0.16$ and $\delta_{CP}/\pi\simeq1.67$.

An approximate solution of $m_\nu^\text{model}=m_\nu$ is
\begin{equation}
 \begin{split}
  &a=0.872-0.277i,\quad B=0.346-0.106i,\quad C=0.748-0.182i, \\
  &d=-0.361-0.098i,\quad Y_1=-0.037-0.271i,\quad Y_2=-0.523+0.036i.
 \end{split}
\label{eq:solution_S3_model2}
\end{equation}
We can tune  $B\to B+\epsilon_B$ with $\epsilon_B=0.033+0.049i$, and we obtain
\begin{equation}
 \begin{split}
  &s_{12}^2=2.94\times10^{-1},\quad s_{13}^2=2.30\times10^{-2},\quad s_{23}^2=5.81\times10^{-1}, \\
  &r=2.93\times10^{-2},\quad\sin\delta_{CP}=0.79,
 \end{split}
\end{equation}
and
\begin{equation}
 \alpha_2/\pi=0.01,\quad\alpha_3/\pi=0.42.
\end{equation}
These observables are consistent with experimental results up to $3\sigma$ except for the $\delta_{CP}$.
The value of $\delta_{CP}$ is out of $3\sigma$ deviation.
The values of Majorana phases are predictions of the model~2.
In this model, we can also tune $a$ or $d$ instead of $B$.

Finally, we fit the values of $Y_i$ by
\begin{equation}
 \frac{Y_2(\tau)}{Y_1(\tau)}\simeq1.917\times e^{-0.479\pi i}.
\end{equation}
By solving this equation, we obtain the solution $\tau=-0.487+0.714i$.
When we change the generation assignment as $L^{(2)}=(L_1,L_3)\to(L_3,L_1)$, we find 
\begin{equation}
 \frac{Y_2(\tau)}{Y_1(\tau)}\simeq0.521\times e^{0.479\pi i},
\end{equation}
and we obtain the solution $\tau=0.480+1.052i$,

\subsubsection{Model 3}
The model~3 predicts 
\begin{equation}
 F^\text{model}\equiv 2(m_\nu^\text{model})_{13}(m_\nu^\text{model})_{23}-(m_\nu^\text{model})_{12}\left[(m_\nu^\text{model})_{33}-(m_\nu^\text{model})_{22}\right]=0.
\label{eq:consistency_S3_IO_model3}
\end{equation}
We define
\begin{equation}
 F_3(\alpha_2;s_{ij},\delta_{CP})=2(m_\nu)_{13}(m_\nu)_{23}-(m_\nu)_{12}\left[(m_\nu)_{33}-(m_\nu)_{22}\right],
\end{equation}
and this is shown in Figure \ref{fig:F6cp} similar to the model~1.
The condition $F_2=0$ is satisfied at $\alpha_2/\pi\simeq0.14$ and $\delta_{CP}/\pi\simeq1.53$.

An approximate solution of $m_\nu^\text{model}=m_\nu$ is
\begin{equation}
 \begin{split}
  &a=0.000+0.188i,\quad B=0.952-0.121i,\quad C=0.458-0.143i, \\
  &d=0.395-0.037i,\quad Y_1=-1.203+0.255i,\quad Y_2=0.289-0.061i.
 \end{split}
\label{eq:solution_S3_model3}
\end{equation}
We can tune $B\to B+\epsilon_B$ with $\epsilon_B=0.001+0.081i$, and we obtain 
\begin{equation}
 \begin{split}
  &s_{12}^2=3.02\times10^{-1},\quad s_{13}^2=2.16\times10^{-2},\quad s_{23}^2=5.89\times10^{-1}, \\
  &r=2.89\times10^{-2},\quad\sin\delta_{CP}=0.84,
 \end{split}
\end{equation}
and
\begin{equation}
 \alpha_2/\pi=-0.03,\quad\alpha_3/\pi=-0.54.
\end{equation}
These observables are consistent with experimental results up to $1\sigma$ except for the $\delta_{CP}$.
The value of $\delta_{CP}$ is out of $3\sigma$ deviation.
The values of Majorana phases are predictions of the model~3.
In this model, we can also tune $C$ or $d$ instead of  $B$.

Finally, we fit the values of $Y_i$ by
\begin{equation}
 \frac{Y_2(\tau)}{Y_1(\tau)}\simeq0.240\times e^{-1.000\pi i}.
\end{equation}
By solving this equation, we obtain the solution $\tau=1.000+1.289i$.
When we change the generation assignment as $L^{(2)}=(L_2,L_3)\to(L_3,L_2)$, we get
\begin{equation}
 \frac{Y_2(\tau)}{Y_1(\tau)}\simeq4.163\times e^{1.000\pi i},
\end{equation}
This equation has no solution.

\section{Conclusion}
\label{sec:Conclusion}

We have studied neutrino mixing in the models with $A_4$ and $S_3$ discrete flavor symmetries.
In our models, couplings are also nontrivial representations under the discrete flavor symmetries, and 
they are modular functions.
In the $A_4$ model, following \cite{Feruglio:2017spp}, 
we assigned the three generations of leptons to the  triplet of $A_4$.
%
%
In this case, the form of the neutrino mass matrix is strongly restricted as (\ref{eq:massmatrix_A4}), and there are no realistic solution.
In the $S_3$ models, we assigned the three generations of leptons to a singlet and a doublet of $S_3$.
In these cases, there are five model parameters except for an overall coefficient in Table~\ref{tab:S3 models}.
It may be easier to fit the experimental data by increasing the number of parameters.
However, neutrino mass matrices in our models have restricted forms, and are written by modular functions.
Thus, it is non-trivial to realize the experimental values 
by many parameters. 
Indeed, there are no solution in the case of normal ordering.
In the case of inverted ordering, we can reconstruct experimental results except for the $\delta_{CP}$ within the $3\sigma$-range in all three models.
Additionally, we can fit the mass matrix by using modular functions in all three models.
Also, we have predictions on the Majorana CP phases.

It would be interesting to study other assignments of leptons in $A_4$ and $S_3$ models such that 
the charged lepton mass matrix is not diagonal and depend on modular functions of $\tau$.
%
%
Also, it would be interesting to extend our analyses to other congruence subgroups, 
e.g. $\Gamma(4) \simeq S_4$ and $\Gamma(5) \simeq A_5$.

\subsection*{Acknowledgement}
The authors would like to thank Y.Takano for useful discussions.
T.~K. was is supported in part by MEXT KAKENHI Grant Number JP17H05395 and 
JSP KAKENHI Grant Number JP26247042.

\appendix 

\section{Modular funcitons}
\label{sec:app-1}

Here, following \cite{Feruglio:2017spp}, we derive modular functions 
with modular weight 2, 
which behave as an $A_4$ triplet and an $S_3$ doublet.

Suppose that the function $f_i(\tau)$ has modular weight $k_i$.
That is, it transforms under the modular transformation (\ref{eq:tau-SL2Z}),
\begin{equation}
f_i(\tau) \rightarrow (c\tau +d)^{k_i}f_i(\tau).
\end{equation}
Then, it is found that 
\begin{equation}
\frac{d}{d\tau}\sum_i \log f_i(\tau) \rightarrow (c\tau +d)^2 \frac{d}{d\tau}\sum_i \log f_i(\tau) + 
c(c \tau +d)\sum_i k_i.
\end{equation}
Thus, $\frac{d}{d\tau}\sum_i \log f_i(\tau)$ is a modular function with the weight 2 if 
\begin{equation}\label{eq:ki=0}
\sum_i k_i = 0.
\end{equation}

We find the following transformation behaviors under $T$,
\begin{eqnarray}
& & \eta(3\tau) \rightarrow e^{i \pi/4} \eta(3\tau), \nonumber \\
& & \eta(\tau/3) \rightarrow \eta((\tau +1)/3), \nonumber \\
& & \eta((\tau + 1)/3) \rightarrow \eta((\tau +2)/3),  \\
& & \eta((\tau + 2)/3) \rightarrow e^{i \pi /12}\eta(\tau /3),  \nonumber 
\end{eqnarray}
and the following transformations under $S$,
\begin{eqnarray}
& & \eta(3\tau) \rightarrow \sqrt{\frac{-i\tau}{3}} \eta(\tau/3), \nonumber \\
& & \eta(\tau/3) \rightarrow \sqrt{-i3\tau} \eta(3\tau), \nonumber \\
& & \eta((\tau + 1)/3) \rightarrow e^{-i\pi/12} \sqrt{-i\tau}\eta((\tau +2)/3),  \\
& & \eta((\tau + 2)/3) \rightarrow e^{i \pi /12}\sqrt{-i\tau}\eta((\tau +1)/3).  \nonumber 
\end{eqnarray}
Using them, we can construct the modular functions with weight 2 by 
\begin{equation}
Y(\alpha,\beta,\gamma,\delta | \tau) =\frac{d}{d\tau} \left( \alpha \log \eta(\tau/3) + \beta \log \eta((\tau + 1)/3) 
+ \gamma \log  \eta((\tau + 2)/3) + \delta \log  \eta(3\tau) \right),
\end{equation}
with $\alpha + \beta + \gamma + \delta = 0$ because of Eq.(\ref{eq:ki=0}).
These functions transform under $S$ and $T$ as 
\begin{eqnarray}
&S:& Y(\alpha,\beta,\gamma,\delta | \tau) \rightarrow \tau^2 Y(\delta, \gamma, \beta, \alpha | \tau), \nonumber \\
&T:& Y(\alpha,\beta,\gamma,\delta | \tau) \rightarrow  Y(\gamma, \alpha, \beta, \delta | \tau).
\end{eqnarray}

Now let us construct an $A_4$ triplet by the modular functions $Y(\alpha,\beta,\gamma,\delta | \tau) $.
We use the $(3 \times 3)$ matrix presentations of $S$ and $T$ as 
\begin{equation}
\rho(S) = \frac 13\left(
\begin{array}{ccc}
-1 & 2 & 2 \\
2 & -1 & 2 \\
2 & 2 & -1
\end{array}\right), \qquad 
\rho(T) = \left(
\begin{array}{ccc}
1 & 0 & 0 \\
0 & \omega & 0 \\
0 & 0 & \omega^2
\end{array} \right) ,
\end{equation}
where $\omega = e^{2\pi i /3}$.
They satisfy 
\begin{equation}
(\rho(S))^2 = {\mathbb I}, \qquad (\rho(S) \rho(T))^3= {\mathbb I}, \qquad (\rho(T))^3={\mathbb I},
\end{equation}
that is, $\Gamma(3) \simeq A_4$.
Using these matrices and $Y(\alpha,\beta,\gamma,\delta | \tau) $, we search an $A_4$ triplet, which satisfy,
\begin{equation}
\left( 
\begin{array}{c}
Y_1(-1/\tau) \\
Y_2(-1/\tau) \\
Y_3(-1/\tau)
\end{array} \right) = \tau^2 \rho(S) \left(
\begin{array}{c}
Y_1(\tau) \\
Y_2(\tau) \\
Y_3(\tau)
\end{array} \right), \qquad 
\left( 
\begin{array}{c}
Y_1(\tau +1) \\
Y_2(\tau +1) \\
Y_3(\tau +1)
\end{array} \right) = \rho(T) \left(
\begin{array}{c}
Y_1(\tau) \\
Y_2(\tau) \\
Y_3(\tau)
\end{array} \right) .
\end{equation}
Their solutions are written by 
\begin{equation}
Y_1(\tau) = 3c Y(1,1,1,-3 | \tau), \qquad 
Y_2(\tau) = -6c Y(1,\omega^2,\omega,0 | \tau), \qquad 
Y_3(\tau) = -6c Y(1,\omega,\omega^2,0 | \tau), 
\end{equation}
up to the constant $c$.
They are explicitly written by use of eta-function as 
\begin{eqnarray} 
Y_1(\tau) &=& \frac{i}{2\pi}\left( \frac{\eta'(\tau/3)}{\eta(\tau/3)}  +\frac{\eta'((\tau +1)/3)}{\eta((\tau+1)/3)}  
+\frac{\eta'((\tau +2)/3)}{\eta((\tau+2)/3)} - \frac{27\eta'(3\tau)}{\eta(3\tau)}  \right), \nonumber \\
Y_2(\tau) &=& \frac{-i}{\pi}\left( \frac{\eta'(\tau/3)}{\eta(\tau/3)}  +\omega^2\frac{\eta'((\tau +1)/3)}{\eta((\tau+1)/3)}  
+\omega \frac{\eta'((\tau +2)/3)}{\eta((\tau+2)/3)}  \right) , \\
Y_3(\tau) &=& \frac{-i}{\pi}\left( \frac{\eta'(\tau/3)}{\eta(\tau/3)}  +\omega\frac{\eta'((\tau +1)/3)}{\eta((\tau+1)/3)}  
+\omega^2 \frac{\eta'((\tau +2)/3)}{\eta((\tau+2)/3)}  \right) , \nonumber
\end{eqnarray}
where we set $c=i/(2 \pi)$.
They can be expanded as 
\begin{eqnarray} 
Y_1(\tau) &=& 1 + 12 q + 36 q^2 + 12 q^3 + \cdots,  \nonumber \\
Y_2(\tau) &=& -6q^{1/3}(1 + 7 q + 8q^2 + \cdots) , \\
Y_3(\tau) &=& -18q^{2/3}(1 + 2 q + 5q^2 + \cdots) . \nonumber
\end{eqnarray}

Similarly, we can construct the modular functions, which behave as an $S_3$ doublet.
Under $T$, we find the following transformation behaviors,
 \begin{eqnarray}
& & \eta(2\tau) \rightarrow e^{i \pi/6} \eta(2\tau), \nonumber \\
& & \eta(\tau/2) \rightarrow \eta((\tau +1)/2),  \\
& & \eta((\tau + 1)/2) \rightarrow e^{i \pi /12}\eta(\tau/2).  \nonumber \\
\end{eqnarray}
Also, $S$ transformation is represented by 
\begin{eqnarray}
& & \eta(2\tau) \rightarrow \sqrt{\frac{-i\tau}{2}} \eta(\tau/2), \nonumber \\
& & \eta(\tau/2) \rightarrow \sqrt{-i3\tau} \eta(2\tau),  \\
& & \eta((\tau + 1)/2) \rightarrow e^{-i\pi/12} \sqrt{-i\tau}\eta((\tau +1)/2). \nonumber  \\
\end{eqnarray}
Then, we consider 
\begin{equation}
Y(\alpha,\beta,\gamma | \tau) =\frac{d}{d\tau} \left( \alpha \log \eta(\tau/2) + \beta \log \eta((\tau + 1)/2) 
+ \gamma \log  \eta(2\tau) \right).
\end{equation}
These functions are the modular functions with the weight 2 if $\alpha + \beta + \gamma =0$.
They transform under $S$ and $T$ as
\begin{eqnarray}
&S:& Y(\alpha,\beta,\gamma | \tau) \rightarrow \tau^2 Y( \gamma, \beta, \alpha | \tau), \nonumber \\
&T:& Y(\alpha,\beta,\gamma| \tau) \rightarrow  Y(\gamma, \alpha, \beta | \tau).
\end{eqnarray}

Using $Y(\alpha,\beta,\gamma | \tau)$, we construct the $S_3$ doublet.
For example, we use the $(2\times 2)$ matrix representations of $S$ and $T$ as 
\begin{equation}
\rho(S) = \frac 12 \left( 
\begin{array}{cc}
-1 & -\sqrt 3 \\
-\sqrt 3 & 1
\end{array}\right), \qquad 
\rho(T) =  \left( 
\begin{array}{cc}
1 & 0\\
0 & -1
\end{array}\right) .
\end{equation}
They satisfy 
\begin{equation}
(\rho(S))^2 = {\mathbb I}, \qquad (\rho(S) \rho(T))^3= {\mathbb I}, \qquad (\rho(T))^2={\mathbb I},
\end{equation}
that is, $\Gamma(3) \simeq S_3$.
Using these matrices and $Y(\alpha,\beta,\gamma | \tau) $, we search an $S_3$ doublet, which satisfy,
\begin{equation}
\left( 
\begin{array}{c}
Y_1(-1/\tau) \\
Y_2(-1/\tau) 
\end{array} \right) = \tau^2 \rho(S) \left(
\begin{array}{c}
Y_1(\tau) \\
Y_2(\tau) 
\end{array} \right), \qquad 
\left( 
\begin{array}{c}
Y_1(\tau +1) \\
Y_2(\tau +1)
\end{array} \right) = \rho(T) \left(
\begin{array}{c}
Y_1(\tau) \\
Y_2(\tau) 
\end{array} \right) .
\end{equation}
Their solutions are written by 
\begin{equation}
Y_1(\tau) = c Y(1,1,-2 | \tau), \qquad 
Y_2(\tau) = \sqrt 3 c Y(1,-1,0 | \tau), 
\label{eq:modular func_S3_doublet}
\end{equation}
up to the constant $c$.
They are explicitly written by use of eta-function as 
\begin{eqnarray} 
Y_1(\tau) &=& \frac{i}{4\pi}\left( \frac{\eta'(\tau/2)}{\eta(\tau/2)}  +\frac{\eta'((\tau +1)/2)}{\eta((\tau+1)/2)}  
- \frac{8\eta'(2\tau)}{\eta(2\tau)}  \right) ,\nonumber \\
Y_2(\tau) &=& \frac{\sqrt{3}i}{4\pi}\left( \frac{\eta'(\tau/2)}{\eta(\tau/2)}  -\frac{\eta'((\tau +1)/2)}{\eta((\tau+1)/2)}   \right)  ,
\end{eqnarray}
where we set $c=i/(2 \pi)$.
Moreover, they can be expanded as  
\begin{eqnarray} 
Y_1(\tau) &=& \frac 18 + 3q + 3q^2 + 12 q^3 + 3q^4 \cdots ,\nonumber \\
Y_2(\tau) &=& \sqrt 3 q^{1/2} (1+ 4 q + 6 q^{2} + 8 q^{3} \cdots  ). 
\end{eqnarray}


\end{document}